# Did a gamma-ray burst initiate the late Ordovician mass extinction?

A.L. Melott[1], B.S. Lieberman[2], C.M. Laird[1], L.D. Martin[3], M.V. Medvedev[1], B.C. Thomas[1], J.K. Cannizzo[4], N. Gehrels[4], & C.H. Jackman[5]

(CORRESPONDING AUTHOR: melott@ku.edu; phone 785-864-3037; fax 785-864-5262)

1. Department of Physics and Astronomy, University of Kansas, Lawrence, KS 66045 USA
2. Departments of Geology and Ecology and Evolutionary Biology, University of Kansas, Lawrence, KS 66045 USA
3. Museum of Natural History, University of Kansas, Lawrence, KS 66045, USA
4. Laboratory for High Energy Astrophysics, NASA Goddard Space Flight Center, Code 661, Greenbelt, MD 20771 USA
5. Laboratory for Atmospheres, NASA Goddard Space Flight Center, Code 916, Greenbelt, MD 20771 USA

***ABSTRACT***
*Gamma-ray bursts (hereafter GRB) produce a flux of radiation detectable across the observable Universe. A GRB within our own galaxy could do considerable damage to the Earth's biosphere; rate estimates suggest that a dangerously near GRB should occur on average two or more times per billion years. At least five times in the history of life, the Earth experienced mass extinctions that eliminated a large percentage of the biota. Many possible causes have been documented, and GRB may also have contributed. The late Ordovician mass extinction approximately 440 million years ago may be at least partly the result of a GRB. A special feature of GRB in terms of terrestrial effects is a nearly impulsive energy input of order 10 s. Due to expected severe depletion of the ozone layer, intense solar ultraviolet radiation would result from a nearby GRB, and some of the patterns of extinction and survivorship at this time may be attributable to elevated levels of UV radiation reaching the Earth. In addition a GRB could trigger the global cooling which occurs at the end of the Ordovician period that follows an interval of relatively warm climate. Intense rapid cooling and glaciation at that time, previously identified as the probable cause of this mass extinction, may have resulted from a GRB.*

**INTRODUCTION**

As mass extinctions have become well-documented, interest in them has grown, partly out of concern for our current environmental situation. Extraterrestrial causes have been seriously considered in recent years. Building on previous studies, we outline the devastating effects a GRB would have on the biosphere and present evidence that a GRB could have contributed to the end Ordovician (~440Mya) mass extinction, which is one of the five great mass extinctions in the history of life. We recognize that the role of a GRB in mediating the extinction is at this time only a "credible working hypothesis" but one worthy of further consideration.

GRB (Mészáros 2001) were initially detected as bursts of X-ray and $\gamma$ radiation, and later determined to lie at cosmological distances, implying very large energies. They are probably associated with some supernovae, producing narrow beams of radiation during gravitational collapse. Some of these beams are visible if they happen to be pointed at the observer. GRB have been, when localized, found to lie in galaxies. It is consistent with observation that our galaxy has hosted GRB, and a rate can be estimated, as well as the probability that any random locale would be irradiated.

It seems likely that a GRB has affected the Earth (Thorsett 1995; Scalo & Wheeler 2002), and that a GRB would have a substantial effect on living organisms. The consequences of a GRB depend on the flux of photons incident on the Earth, not the intrinsic total energy of the GRB. For most purposes, beaming is irrelevant, as the actual GRB rate inferred from data varies inversely with the beaming angle, leaving the observed rate constant. Due to small number statistics, lack of knowledge of the recent GRB rate is a serious source of uncertainty in estimating the role they may have played in mass extinctions. Most GRB are at high redshift, and the cosmological evolution of the rate is poorly constrained (Weinberg et al., 2001). However, even the most conservative rate estimates indicate that a catastrophic event within the geological record is likely.

**THE LOCAL RATE OF "DANGEROUS" GRB EVENTS**

We discuss the threat by estimating the probable interval between "dangerous" GRB events. Of course, the threat continuously varies with distance. For definiteness, we estimate the rate of events at low redshift (most of the geological record lies within the time corresponding to cosmologically low redshifts). We estimate the distance and power of the nearest probable event within the disk of our galaxy (which is sufficiently thin to treat as two-dimensional for this purpose). As we shall see, that a danger exists has a high certainty. For reasons that will emerge from our later atmospheric discussion, we estimate that a typical GRB pointed at the Earth from 3 kpc (or about 10,000 light years) or closer constitutes a serious threat to the biosphere.

There are various ways to estimate the GRB frequency. One approach is to use best fits to an estimate of the GRB rate at low redshift, about 0.44 $Gpc^{-3}$ $y^{-1}$ (Guetta et al. 2003). This can be compared with the number density of typical galaxies, about $3 \times 10^{-3}$ $Mpc^{-3}$ (using a fit to the luminosity function in Loveday et al (1992), and assuming h=0.7). This gives an event rate of $1.5 \times 10^{-7}$ $gal^{-1}$ $y^{-1}$; since 3 kpc is about one-fifth the radius of our galaxy, we would expect a dangerous event rate of $6 \times 10^{-9}$ $y^{-1}$, in other words about 170

My between events within 3 kpc. This is a crude method, if only because the galaxy does not have a definite boundary.

A cleaner approach, following Scalo and Wheeler (2002) is to scale from blue luminosity density. Blue luminosity is produced by the large, hot stars with short lifetimes that are thought to be the precursors of most GRB. They present results based on various assumed rates of GRB frequency evolution—the rate of GRB is thought to have been larger in the past, consistent with scaling with the star formation rate. We cite estimates for the strong evolution case. This produces the smallest number of GRB within "recent" cosmological time, when the five largest mass extinctions occurred. Current research (e.g. Kewley et al. 2004) favors older estimates of strong evolution in the star formation rate, when reddening is corrected. In this case the typical distance to the nearest GRB irradiating a locale within our galaxy would be 1.9 kpc $t_G^{-1/2}$, where $t_G$ is the time interval in Gy. As we shall see, the greatest threat appears to come from ozone depletion resulting in increased solar UV reaching the Earth's surface. The ozone shield is thought to have been significant for about 1 Gy, so we shall use this timescale. We expect the nearest burst irradiating the Earth to have been at a distance of about 2 kpc over the last Gy. This is of course only an average. This method may bias the result downward somewhat, because there is evidence that GRBs preferentially go off in regions of low galaxy density, rather like the one in which our galaxy is situated (Bornancini et al. 2004). Thus, the rate in the Milky way might be somewhat higher than the global average. This is not expected to be a large effect.

Another method is to use the fit of Guetta et. al (2003) to the low redshift effective GRB rate, again 0.44 Gpc$^{-3}$ y$^{-1}$. Using the universal blue luminosity density of 6.0 x 10$^{43}$ W Gpc$^{-3}$ (for h=0.7; Zucca et al 1997) and the blue luminosity surface density of our galaxy 8.0 x 10$^{27}$ W pc$^{-2}$ (Binney and Merrifield 1998) we estimate a rate of 5.8 x 10$^{-11}$ y$^{-1}$ kpc$^{-2}$. Using the Poisson process model, the estimated distances to nearest events are only 10% greater than those of Scalo and Wheeler (2002). This is excellent agreement, since their estimate is based on fits to evolution of the star formation rate, and the latter estimate is based on the independent Guetta et al. (2003) estimate of the GRB evolution rate.

The latest results (Firmani et al. 2004), scaled from the star formation rate in the Milky Way galaxy, suggest a higher frequency, about 1.5 x 10$^{-7}$ y$^{-1}$ in our galaxy (pointed at us). Again assuming a 15 kpc nominal radius for our galaxy, the Poisson model would place the typical nearest burst much closer, at about 1 kpc $t_G^{-1/2}$. Firmani et al. represent an extreme assumption, which is however not inconsistent with GRB evolution data (e.g. Weinberg et al. 2001).

Of course, the Poisson process model ignores local rate variation, such as enhanced star formation in spiral arms. But given that we are averaging over long times, in which our solar system has migrated in and out of such regions, as well as the possibility of an event which might have occurred within a large fraction of the surface area of this galaxy, the approach should be valid for mean rates. Most GRB are probably beamed, with γ-rays confined to a narrow cone. However, these estimates give the probability based on the observed flux; if they are beamed, events occur more often, with only a small fraction aimed at a given observer.

GRB have a wide range of duration and "isotropic equivalent energy"—the energy of the GRB if the observed flux is assumed to be radiated equally in all directions. Guetta et al. (2003) estimate a typical isotropic equivalent luminosity of $4.4 \times 10^{44}$ W at low redshift from an analysis of BATSE data and the observed redshift distribution of events for which this information is available. Lloyd-Rooning et al (2000) deduce a slightly larger number, about $6 \times 10^{44}$ W from a statistical analysis of 220 BATSE GRB with redshifts estimated from a luminosity-variability relation. We use average power $5 \times 10^{44}$ W for purposes of this discussion.

It can be shown that due to the shallow slope of the GRB luminosity function, the rarer, more distant, and luminous GRB offer an additional threat of comparable magnitude. So far we have presented estimates for a local event given a "typical GRB". However, GRB show large diversity in total isotropic luminosity ranging from about $10^{43}$ to $10^{47}$ W. Here we estimate the contribution of more energetic but less frequent events. Yonetoku et al. (2003) exploited the correlation (Amati et al. 2002) between the isotropic energy and the peak energy of the GRB spectrum to derive the luminosities of 684 GRB. The derived cumulative luminosity function $N(>L)$, i.e., the number of GRB with luminosity greater then L, at redshift zero (i.e., the luminosity evolution with redshift being removed), is approximated at luminosities greater than $10^{45}$ W as

$N(>L) \sim (L/10^{47} \text{ W})^{-1.4}$

To produce the same energy flux at the Earth a more energetic GRB may be seen from a larger distance:

$F_{obs} = L/R^2 \sim$ const.

This increases the chance probability of such a GRB being within the distance R from us:

$P = V_{GRB}/V_{Galaxy} \sim R^2$,

where $V_{GRB}$ and $V_{Galaxy}$ are the volume of the Galaxy within the distance R from a GRB and the total volume of the Galaxy, respectively. Note that with R being of order or greater than 3 kpc, the galactic disk may be treated as two-dimensional. Note also that because of large dust opacity around the Galactic center, $V_{GRB}$ cannot be greater than about 50% of $V_{Galaxy}$. Normalizing the rate of dangerous events to $R_t$, the rate of typical GRB, we estimate the rate of dangerous events as a function of luminosity to be $\sim R_t (L/10^{45} W)^{-0.4}$. As one can see, the dependence on the luminosity is very shallow, which means that large but rare events are almost as dangerous as more frequent and less energetic typical GRB. Ultimately, this increases the total rate of potentially dangerous GRB by a factor of two.

There also exists a category of short-burst GRB, about which much less is known. They have much harder photon spectra but also a lower rate and lower total energy. We do not include them in our estimate of the hazard to biospheres.

The most probable equivalent distance to the nearest event in the last Gy could range from 0.7 to 2 kpc, depending on which of the event rate estimates we adopt and whether or not we include the effect of more distant powerful GRBs. Two events GRB1 and GRB2 bracket the range. GRB1 and GRB2 both have equivalent power of 5 x $10^{44}$ W for 10s. GRB1 is at 2 kpc and GRB2 is at 700 pc. GRB1 uses the low end of the rate estimates; GRB2 uses the high end and includes as a correction the contribution of the more luminous, distant events. In both cases we assume a burst of 10s. From GRB1 the upper atmosphere of the Earth is irradiated by about 10 kW/m², and in GRB2 it is about 80 kW/m² in X-rays and γ-rays.

It has been suggested that copious quantities of ultra high-energy cosmic rays may be emitted by GRB (eg Wick et al. 2003, Waxman 2004). If this were true, cosmic rays above $10^{18}$ eV would travel from our fiducial bursts with so little deflection as to arrive within the beam of the burst photons. However, in the absence of any direct detection of such cosmic rays, or the neutrinos that would be expected to accompany them (Kravchenko et al 2003), we will not as yet consider this additional stress on the biosphere.

**ATMOSPHERIC AND BIOLOGICAL EFFECTS OF GRB**

The flux of order 10 kW/m² from GRB1 is at first sight not unusual, since it is about three times the intensity of sunlight at the radius of the Earth's orbit. However, its effects are far different. Photon energies are typically hundreds of keV, in the X-ray and γ-ray region, rather than eV, the visible light region. The mean free path of such photons in the atmosphere is quite short, and most of the energy is dumped in the stratosphere where ozone concentrations are highest (Gehrels et al. 2003). Moreover, the effects are quite different: each photon is able to produce many ionizations or dissociations of molecules, about one ionization per 35 eV, as used in Gehrels et al (2003). Smith et al (2003) have computed the propagation of such photons in Earthlike planetary atmospheres. About 0.002 of the energy reaches the surface, primarily in the form of dangerous UVB (280-320 nm) photons. The short burst UV power of 20 W/m$^2$ from GRB1 is about five times the typical UVB flux at the Earth's surface on a bright sunny day, but is not likely to be a major effect. On the other hand, 160 W/m$^2$ from GRB2 reaching the ground would likely be quite damaging, even in a short period. So, the danger to one side of the Earth in the initial burst depends critically upon the proximity of the GRB. Most of the burst energy incident upon the atmosphere goes into ionization and dissociation of molecules. As we shall see, this has much more disruptive long-term effects.

If ultra-high energy cosmic rays (mostly protons) accompanied the GRB, they could irradiate the surface with muons, secondaries produced by interaction with the atmosphere, while producing radionuclides by spallation. However, as discussed earlier the cosmic ray contribution is speculative. Most cosmic rays will scatter and add a contribution comparable to the background from supernovae. Most GRB photons are insufficiently energetic to induce any nuclear reactions. Thus, the instantaneous biotic effects will be UV-dominated, and for our minimal GRB, modest (compared to the long-term effects) and confined to the facing side of the Earth. However, we identify as a major uncertainty the controversy over GRB cosmic ray production, that could in principle greatly exacerbate the effects (Waxman 2004; Wick et al.2003).

GRB may be approximately as disruptive as supernovae, on average, or considerably more so, depending on assumptions. Gehrels et al. (2003) computations show that a supernova at a distance of 8 pc is likely to happen about 1.5 $Gy^{-1}$, close to the assumed frequency of the GRB events we consider. It would irradiate the atmosphere with about 25 $kJ/m^2$, which is one-fourth that of GRB1. The fluence in excess of 1 $MJ/m^2$ from GRB2 is far in excess of any likely result from a supernova. Moreover, the supernova irradiation would be spread out over a period of many months rather than seconds. Any processes that act to restore the former equilibrium, such as the destruction of oxides of nitrogen, will reduce the impact of the supernova, since the synthesis of these compounds is essentially instantaneous. Previous computations (Laird et al. 1997) show that the timescale for decay of oxides of nitrogen and the timescale for recovery of the ozone layer, once these destructive catalysts are removed, is in the order of several years. The slow steady irradiation from a supernova must compete with decays, but the GRB irradiation accumulates rapidly. As we now discuss, the $NO_x$ production should be somewhat self-limiting, partially due to recombination of N atoms with NO molecules to form $N_2$.

The $NO_x$ (N, NO, $NO_2$) production is somewhat difficult to estimate. The nitrogen atoms produced through the dissociation of $N_2$ by the gamma rays and their associated secondary electrons create NO through the reaction $N + O_2 \longrightarrow NO + O$ at a rate of $k_1=1.5 \times 10^{-11} \exp(-3600/T)$ (Sander et al., 2003) and destroy NO through the reaction $N + NO \longrightarrow N_2 + O$ at a rate of $k_2=2.1 \times 10^{-11} \exp(100/T)$ [Sander et al., 2003]. The highest energy gamma rays can penetrate to about the middle of the stratosphere, where they have the most influence on total ozone. Our typical GRB photons penetrate to near the middle of the stratosphere at 10 hPa (~30 km); most of the ozone is near this altitude. With an assumed temperature of 220K, the timescale for reaction of N with $O_2$ is about 9 s [$k_1 = 1.2 \times 10^{-18}$ cm$^3$ s$^{-1}$] and for reaction of N with NO is about 25 s [$k_2 = 3.3 \times 10^{-11}$ cm$^3$ s$^{-1}$], assuming an NO mixing ratio of 2.5 ppbv. The NOx will thus increase substantially in the middle stratosphere before being self-limited through the second reaction. If the region is heated by about 10K, as seems reasonable, the creation of NO will go slightly faster and its destruction slightly slower.

There is an additional possible reaction that may limit $NO_x$ production. This is N + N --> N2 + γ, which could in principle destroy N atoms before they can go on to make $NO_x$. This reaction has a very limited phase space and is highly unlikely due to the simultaneous requirement of momentum and energy conservation. A more likely reaction is the three-body reaction N + N + M --> N2 + M + γ, where M is an arbitrary reactant. However, this is likely to be a small effect, and rate-limiting for very large $NO_x$ production levels only. The extreme GRB2 produces about $5 \times 10^{13}$ ionizations/cm$^3$. At 10 hPa, the total number density is about $3 \times 10^{17}$ /cm$^3$. The $N_2$ abundance would thus be about $2.3 \times 10^{17}$ /cm$^3$ and $O_2$ abundance would be about $6.3 \times 10^{16}$ /cm$^3$. Assuming that we produce about 1.25 N atoms / ionization, then there would be $6.2 \times 10^{13}$ N atoms /cm$^3$ produced by GRB2. The ratio of N to $N_2$ would be $2.7 \times 10^{-4}$; the ratio of N to $O_2$ would be $9.8 \times 10^{-4}$; and the ratio of N to M (total number density) would be $2.1 \times 10^{-4}$. Thus, the N abundance compared with $O_2$ or $N_2$ is quite small; they would be even smaller for GRB1. It would thus seem logical that there is more probability of N reacting with $O_2$ than N reacting with N (especially in a three body reaction). It therefore seems very probable that a GRB will create NO, even though the production is self-limited by the $N + NO \rightarrow N_2 + O$ reaction. This rate limitation should only come into play after rather high NO densities are reached.

Rusch et al. (1981) discuss the efficiency for production of $NO_x$ during an extreme particle precipitation period, the August 1972 solar proton event, and show that the efficiency for $NO_x$ production is dependent on altitude and duration of the event. The reaction rates $k_1$ and $k_2$ for $N + O_2$ and $N + NO$, respectively, assume that nitrogen atoms are in their ground state, $N(^4S)$. Rusch et al. (1981) present model simulations with the branching ration for production of nitrogen atoms in the excited state, $N(^2D)$, from 0.5 to 1.0. They conclude that a branching ratio of 0.8 for production of $N(^2D)$ appears to produce an ozone destruction most compatible with the measurements. The reaction rate of $N(^2D) + O_2 \rightarrow NO + O$ [$k_3 = 5.3 \times 10^{-12}$ cm$^3$ s$^{-1}$, from Rees (1989)] is much faster than $k_1$, given above. The self-limiting $N(^2D) + NO \rightarrow N_2 + O$ [$k_4 = 7 \times 10^{-11}$ cm$^3$ s$^{-1}$ from Rees (1989)] reaction rate is similar in value to reaction rate $k_2$. This discussion suggests that $NO_x$ will be increased by a GRB, possibly in a substantial way, and will cause a depletion of ozone. It is possible that other reactions need to be considered during the 10 second duration of a GRB, in order to perform a reliable computation of the ozone depletion. A detailed computation of the ozone impact is outside the scope of this paper and is under careful investigation.

Long-term effects of GRB would spread around the Earth (by prevailing winds and diffusion processes) and include ozone layer depletion, acid rain, and global cooling. The effects of a nearly polar burst would never fully affect both hemispheres, because the timescale for diffusion into the opposite hemisphere is in the order of a few years, equaling the decay time of the catalytic products. However, due to the solid angle subtended, a purely polar burst is highly improbable. Bursts as high as 60 degrees latitude that spill significantly into the opposite hemisphere have considerable impact on the opposite hemisphere extending all the way to the pole in a few months, depending on time of year. On the other hand, longitudinal diffusion around the Earth takes only a week or two.

The chemical effects result primarily from ionization and dissociation of molecules of $N_2$ and $O_2$, the dominant constituents of the atmosphere. The resulting highly reactive products of $N_2$ dissociation are various oxides of nitrogen. Nitric acid comparable to or exceeding anthropogenic levels is a possible product (Thorsett 1995). Global cooling may occur (Reid 1978) from the absorption of visible light by $NO_2$. On the other hand, very substantial ozone depletion will result from $NO_x$ constituents, which catalyze conversion (Gehrels et al. 2003) of ozone to oxygen molecules. $NO_x$ species persist several years in the atmosphere (Jackman et al. 2000), so there should be a comparable timescale for ozone depletion. Ozone now absorbs about 98% of biologically damaging solar UV radiation before it reaches the surface, so the depletion of the ozone layer would have serious consequences. A GRB may have paradoxically produced darkened skies and heightened UV radiation.

Our minimal GRB1 event would produce about $6 \times 10^{12}$ ionizations/cm$^3$, mostly in the stratosphere, using 35 eV per ionization, as in Gehrels et al. (2003). GRB2 would produce about $5 \times 10^{13}$ ionizations/cm$^3$. In comparison, the largest solar proton event in the last 40 years produced of order $1.7 \times 10^9$ ionizations/cm$^3$, also in the stratosphere (Jackman et al 1995). This event was computed to have produced a maximum total ozone depletion of about 4%. Gehrels et al. (2003) simulations showed significant ozone depletion for a supernova with fluence four times less than that of GRB1, albeit spread over many months. A linear extrapolation is not valid, but it is possible that a GRB could produce significant ozone depletion. The short timescale and high intensity of the burst is a significant challenge for evaluating the atmospheric changes, beyond the scope of this paper. Pending the results of these computations, we regard 3 kpc as a reasonable estimate of the proximity at which a GRB becomes disruptive. Such an event has twice the fluence of the likely nearest supernova and many orders of magnitude more than a solar flare which produced significant ozone depletion. With the conservative assumptions leading to GRB1, such an event is likely every 500 My. Even with $NO_x$-limiting recombination processes, it seems likely to have significant damaging effects.

Even modest (10-30%) increases in UV flux, particularly around 300 nm, can be lethal to a variety of organisms (Kiesecker et al. 2001; Hader et al. 2003; see also Cockell & Blaustein 2000), including phytoplankton the basis of the marine food chain as well as oxygen production. UV is attenuated by water, though the precise absorption is heavily dependent upon particulates and dissolved organics (Hader et al. 2003). Penetration depths vary from meters to tens of meters. As would be expected, UV effects on microorganisms have been found to decrease with water depth (Sommer et al. 1999).

Given the total number of ionizations expected from our GRB, of order $10^{18-19}/cm^2$, it seems possible that the nitrate deposition could be substantial. From the opacity of the $NO_2$, (a brown gas) there is a possible significant reduction in visible light as a consequence of this event. The reduction in global temperatures resulting from this may have possibly initiated the Ordovician ice age, which came on suddenly in the middle of a period of high climate stability.

**ASPECTS OF THE LATE ORDOVICIAN MASS EXTINCTION POTENTIALLY COMPATIBLE WITH A GRB**

The late Ordovician is one of the largest mass extinctions in terms of its scale and scope. It appears to comprise two large, abrupt extinction events, separated by 0.5-2 My (Brenchley et al. 1994, 1995, 2003; Hallam and Wignall 1997). All major marine invertebrate groups show high rates of extinction during this interval. The late Ordovician mass extinction has been related to a period of rapid cooling followed by a period of rapid global warming (Brenchley et al. 1994, 1995, 2003; Orth et al. 1986). Brenchley (1984), Brenchley et al. (1991, 1994, 1995, 2003), Briggs et al. (1988), Owen et al. (1991), Fortey and Cocks (2003), Sheehan (2001), and others, have convincingly demonstrated how global cooling and warming events could have led to a mass extinction, and we do not dispute the contribution of global cooling to this extinction.

Prior to the late Ordovician cooling, temperatures were relatively warm and it is the suddenness of the climate changes and the elimination of habitats due to sea-level fall that are believed to have precipitated the extinctions (Brenchley et al. 1991, 1994, 1995, 2003; Melchin & Mitchell 1991; Owen et al. 1991; Sheehan 2001). We emphasize that there is evidence for a possible link between GRB and global cooling. GRB produce opaque nitrogen dioxide in Earth's atmosphere, providing a mechanism for global cooling. Berner (1994), Brenchley et al. (1994), and Sheehan (2001) remarked that it is a puzzle that a glaciation developed in the late Ordovician given warm climatic conditions in the middle Ordovician, though Brenchley et al. (1995, 2003) did provide an intriguing explanation based on an increase in the rate of organic carbon burial. It is noteworthy that palaeoclimate models (Herrmann & Patzkowsky 2002; Herrmann et al. 2003) have shown that late Ordovician glaciation could not have occurred without some external forcing mechanism; one such impulse could be reduced solar insolation precipitated by a GRB.

The late Ordovician has been studied comprehensively (e.g., Brenchley et al. 1991, 1994, 1995, 2003; Melchin & Mitchell 1991; Owen et al. 1991; Droser et al. 1996; Adrain et al. 1998; Anstey et al. 2003; Fortey & Cocks 2003) yet generalizations about which species go extinct and why are difficult. (Indeed, this is also true of all the other known mass extinctions.) There may be some patterns of extinction, however, compatible with an initial burst of UV radiation followed by a greatly reduced ozone layer, subsequent relatively rapid global cooling followed by global warming. The oxygen level of the atmosphere in the Ordovician was not greatly different from current levels (Berner et al.

2003), and an ozone shield was in place. Its strong reduction would almost certainly produce catastrophic consequences associated first with more UV radiation reaching the Earth's surface similar to those observed in modern organisms (Kiesecker et al. 2001; Hader et al. 2003).

Notably, one would predict planktonic organisms or larvae would more likely be affected than benthic organisms because the former are less shielded from radiation (unless they occupied deep water). Similarly, epifaunal organisms would be more affected than infaunal organisms buried and shielded in sediments. Both of these patterns may exist in the late Ordovician. Further, Brenchley et al. (2003) suggest that some extinctions may have begun before the episode of global cooling, indicating the possibility of a precursor event to the glaciation, and supporting the notion that global cooling alone may not be sufficient to explain the mass extinction.

One factor that correlates well with increased likelihood of extinction in trilobites, one of the important groups in the Ordovician (e.g., Briggs et al. 1988), is the amount of time a typical organism spent in the water column, especially as a larva. Chatterton & Speyer (1989) quantified the proportion of genera of trilobites with different larval types that went extinct at the end of the Ordovician. Trilobites inferred to have had a planktonic larval phase are proportionately 1.5 times more likely to go extinct than species with a benthic larval phase. In fact, it appears to not only have been whether the larvae were planktonic but the amount of time the larvae spent in the plankton: longer planktonic larval phases, when such can be inferred, are associated with increased extinction probability relative to that of shorter planktonic larval phases.  Trilobite species whose larvae floated in the water column as opposed to those that crawled and were partly protected in sediments would be expected to be much more vulnerable to the initial affects of a GRB, including the elimination of the ozone layer leading to exposure to high levels of UV radiation.

The jury is still out on whether organisms with planktonic larvae suffered high extinction rates during any of the four other mass extinctions known to have occurred in the last 500 My.  The Cretaceous-Tertiary mass extinction, attributed to bolide impact and when the dinosaurs and numerous marine organisms went extinct, is particularly well characterized.  For marine organisms from the Cretaceous period that are well known and have been intensively studied the difference between extinction in non-planktonic and planktonic species is not statistically significant (Smith & Jeffery 1998; also see Valentine & Jablonski, 1986).  Clearly more data are needed on all of the mass extinctions, including the late Ordovician, to pin down this correlation, but analyses available now are at least suggestive.

The differential extinction not only of species with a planktonic larval type, but also planktonic organisms in general, may hold for a variety of taxa in the late Ordovician, though it is hard to draw precise conclusions (Sheehan 2001).  Several groups of marine organisms with a planktonic lifestyle more exposed to UV radiation than groups that lived in the benthos, suffered severely during the late Ordovician.  These groups include graptolites, acritarchs, and cephalopods (Martin, 1985; McKinney 1985; Colbath 1986;

Melchin & Mitchell 1991; Sheehan 2001). Most trilobites, irrespective of their larval phase, are inferred to have occupied a benthic habit as adults; however, there were some Ordovician species whose adults lived in the plankton, and these species were completely eliminated in the mass extinction (Chatterton & Speyer 1989; Owen et al. 1991). Brenchley et al. (2003) have also shown that typically organisms that dwelled in the plankton were affected before benthic organisms during the mass extinction, and species dwelling in shallow water were more likely to go extinct than species dwelling in deep water (also see Brenchley et al., 1995 and Sheehan 2001). Benthic and deep water organisms would typically be more shielded from UV radiation than planktonic and shallow water organisms.

Also interesting is the pattern of extinction in bivalves during the late Ordovician crisis, with burrowing infaunal forms that would be shielded from UV radiation more likely to survive than epifaunal forms living on the surface that were not shielded (Kriz 1984). Finally, the surviving faunas that populated the subsequent Silurian marine world appear to have come from areas of deeper water and/or higher latitude (Melchin & Mitchell 1991; Sheehan 2001).

**CONCLUSIONS**

Radiation from a GRB ionizes and dissociates molecules in the atmosphere, with a burst of UV radiation reaching the ground. Further, this triggers depletion of the ozone layer leading to greatly increased solar UV reaching the surface, and also causing other affronts to life including acid rain possibly followed by global cooling; the result: a one, two punch for life on the planet. Given estimated GRB rates, it is probable that at least one significant event occurred in the last billion years. We present as a credible working hypothesis that a GRB could have contributed to the late Ordovician extinction first by subjecting the Earth's biota to elevated levels of UV radiation, partially due to the initial burst but primarily by depleting the ozone layer, and then by precipitating relatively rapid global cooling which was followed by rapid global warming. At present we see reasons for associating a GRB with the Ordovician mass extinction but clearly additional tests are required. Additional atmospheric chemistry modeling is needed to pin down the radiation fluence needed to trigger given ozone depletion, nitric acid rain, and global cooling levels. Given the uncertainty in the evolution of the GRB rate, it is possible that they were involved in other mass extinctions, or that many much more distant GRBs could have smaller effects (Scalo & Wheeler 2002). A major challenge for astrophysics is to evaluate the likely flux and spectrum of cosmic rays accompanying a GRB.

**ACKNOWLEDGEMENTS**

A.L.M. and B.S.L. acknowledge support from the U.S. National Science Foundation; B.S.L. acknowledges support from a Self Faculty Award; B.C.T. acknowledges support from a Self Graduate Fellowship; and A.L.M. and B.C.T. acknowledge support from NASA and the Graduate Research Fund at the University of Kansas. R. Kaesler, D. Smith, J. Scalo, S. Thorsett, and A. Miller provided helpful comments on earlier versions of this paper. Referee J.C. Wheeler made a number of extremely helpful suggestions which improved presentation.